\documentclass[aps,prd,groupedaddress,nofootinbib]{revtex4-1}

\usepackage{graphicx}
 \usepackage{amsmath}
 \usepackage{color}
 \usepackage{float}
 \usepackage{tikz}
\restylefloat{table}
\newcommand{\as}{\alpha_s}
\newcommand{\mupi}{\mu_\pi^2}
\newcommand{\mug}{\mu_G^2}
\newcommand{\rd}{\rho_D^3}
\newcommand{\rls}{\rho_{LS}^3}

\newcommand{\y}{{\hat q^2} }

\long\def\symbolfootnote[#1]#2{\begingroup%
\def\thefootnote{\fnsymbol{footnote}}\footnote[#1]{#2}\endgroup}

\def \be{\begin{equation}}
\def \ee{\end{equation}}
\newcommand{\bea}{\begin{eqnarray}}
\newcommand{\eea}{\end{eqnarray}}

\newcommand{\qz}{q_0}
\newcommand{\qq}{q^2}
\newcommand{\el}{E_\ell}

\begin{document}

\title{NNVub: a Neural Network Approach to $B\to X_u \ell \nu$ }



\author{Paolo Gambino$^a$}
\author{Kristopher J. Healey$^a$}
\author{Cristina Mondino$^{a,b}$}
\affiliation{$^a$ Dip. di Fisica, Universit\`a di Torino \& INFN, Torino, 10125 Torino, Italy}
\affiliation{$^b$ 
Center for Cosmology and Particle Physics, Physics Dept., New York University, New York, NY 10003, USA}

\date{\today}

\begin{abstract}
We use artificial neural networks to parameterize the shape functions in inclusive semileptonic $B$ decays without charm. Our approach avoids the adoption of functional form models and allows for a straightforward implementation of all experimental and theoretical constraints on the shape functions.
The results are used to extract $|V_{ub}|$ in the GGOU framework and compared with the original GGOU paper and  the latest HFAG results, finding
good agreement  in both cases. The possible  
impact of future Belle-II data on the $M_X$ distribution is also discussed.
 
 \end{abstract}


\maketitle
\section{Introduction}
The precise determination of the CKM matrix element $V_{ub}$ remains an important goal in flavour physics, instrumental in performing stringent tests of the CKM matrix unitarity, see 
\cite{Bevan:2014iga} for a review.
 $|V_{ub}|$ can be extracted from $b\to u \ell\nu$ decays, and in 
in particular from $B\to\pi \ell\nu$, using lattice QCD \cite{Lattice:2015tia,Flynn:2015mha} or light-cone sum rules \cite{Imsong:2014oqa,Bharucha:2012wy} calculations of the relevant form factors. The other exclusive channels  $\Lambda_b\to p \ell\nu$ \cite{Detmold:2015aaa} and $B\to \pi\pi\ell\nu$ \cite{Faller:2013dwa,Kang:2013jaa} are also actively pursued.
 The inclusive determination relies instead on a local Operator Product Expansion (OPE) \cite{Chay:1990da,Bigi:1992su, Bigi:1993fe,Blok:1993va,Manohar:1993qn} which has been successfully applied to $B\to X_c \ell \nu$, see \cite{Alberti:2014yda} for the state of the art.
In the charmless case $B\to X_u \ell \nu$ the convergence of the local OPE is hampered by the experimental cuts that are generally applied to suppress the charm background and that introduce a sensitivity to the Fermi motion of the $b$ quark inside the $B$ meson\footnote{Some of the recent experimental analyses employs sufficiently low cuts to capture up to 90\% of the events, justifying the use of the local OPE. However, these analyses heavily depend on the background subtraction and on the theoretical description of the signal in the shape-function region, whose understanding remains central for an accurate determination of $|V_{ub}|$ from semileptonic $B$ decays.}.
The well-known solution \cite{Neubert:1993ch,Bigi:1993ex} is to introduce a distribution function or {\it Shape Function} (SF) whose moments are dictated by the local OPE. The SF is actually the parton distribution function of the $b$ quark in the meson.
 Effects formally suppressed by $\Lambda_{\rm QCD}/m_b$ are also important and lead to the emergence of additional, largely unknown, shape functions \cite{Bauer:2001mh,Leibovich:2002ys,Bosch:2004cb}.
 The present HFAG inclusive $|V_{ub}|$ average \cite{hfag} is
based on different approaches \cite{Andersen:2005mj,Gambino:2007rp,Lange:2005yw,Aglietti:2007ik}. In the GGOU approach \cite{Gambino:2007rp} it reads
\be
|V_{ub}|_{incl}=(4.51\pm 0.16^{+0.12}_{-0.15})\times 10^{-4}
\ee
and very close values are found with the other methods. The high luminosity expected at Belle-II, together with precise lattice determinations of $f_B$,
will also allow for an accurate determination of $|V_{ub}|$ from decay $B\to \tau \bar\nu$. 

The inclusive and exclusive determinations of $|V_{ub}|$ have been in conflict for a long time \cite{pdg}. Although the latest lattice calculations \cite{Lattice:2015tia,Flynn:2015mha} imply a somewhat larger $|V_{ub}|$ than in the past, the discrepancy is still at the level of almost 3$\sigma$ and calls for further scrutiny of all aspects of these determinations. In the case of the inclusive one, the major open problems are 
$i)$ the limited knowledge of leading and subleading SFs; $ii)$ the
non-perturbative effects in the high-$q^2$ region\footnote{Weak Annihilation contributions are strongly constrained by semileptonic charm decays \cite{WA}.}; $iii)$ the potential role of higher order perturbative effects. The SFs uncertainty, in particular, has been estimated to affect $|V_{ub}|$ only at the level of a few percent \cite{Gambino:2007rp,Lange:2005yw}. However, these analyses were performed assuming a set of two-parameters functional forms, and it is unclear to what extent the chosen set is representative of the available functional space, and whether the estimated uncertainty really reflects the limited knowledge of the SFs. This point was emphasized in \cite{Ligeti:2008ac}, where a different strategy was also proposed, based on the expansion of the leading SF in a basis of orthogonal functions, fitting its coefficients to the $B\to X_s \gamma$ spectrum, and on the modelling of the subleading SFs.

In this paper we introduce a new method based on the Monte Carlo approach,
with neural networks used as unbiased interpolants for the SFs, in a way similar to what the NNPDF Collaboration do in fitting for Parton Distribution Functions \cite{Ball:2008by} and DIS structure functions \cite{Forte:2002fg}.\footnote{The possible use of neural networks to parameterize the SF in semileptonic $B$ decays has been mentioned in Ref.~\cite{Rojo:2006kr}. }
 There are of course several differences with PDF fits, most notably that we parameterize functions of two parameters, and that
direct experimental information on the SFs is presently rather scarse: we only have
measurements of the photon spectrum of $B\to X_s \gamma$ above $\sim1.9$ GeV 
\cite{bsgamma} and OPE constraints on the first moments of the SFs. However, the photon spectrum in inclusive radiative decays does not provide direct information on the SFs that appear in the semileptonic decay beyond leading order in $1/m_b$, and the moments do not constrain the functional form much, as we will see.

While there are several methods to determine a probability distribution function like the SF from its first moments (the {\it truncated moment problem} or Stieltjes moment problem in mathematical analysis), see {\it e.g}.\ \cite{math}, the high flexibility of neural networks allows for the straightforward inclusion of additional constraints, such as 
 the kinematic distributions of $B\to X_u \ell \nu$ which will
be measured with good accuracy at the upcoming Belle-II experiment \cite{future}. The measurement of the $M_X$ or $E_\ell$ shapes, for instance, will contribute useful information on the SFs and in turn reduce the SF uncertainty in the $|V_{ub}|$ extraction.
 
In the following we adopt the GGOU approach, where inclusive semileptonic decays without charm are described in terms of three $q^2$-dependent SFs, whose first moments are known
from the local OPE. This is the minimal set of SFs, and in this approach they are not split into a leading and several subleading SFs. The kinematic distributions accessible at Belle-II will therefore probe some of their combinations. The neural network method presented here provides a simple way to determine $|V_{ub}|$ taking into account all the constraints on the SFs, including all uncertainties and correlations properly. The SFs appearing in $B\to X_s \gamma$ and $B\to X_s \ell^+\ell^-$ can be treated with the same formalism.

The paper is organized as follows. In the next Section we recall the elements of the GGOU approach which are relevant for our topic. In Section III we discuss artificial neural networks and the way we apply them to our problem. Section IV presents our results 
on the resulting SF uncertainty for $|V_{ub}|$, a new extraction of $|V_{ub}|$ from present data, and a preliminary discussion of the improvements possible using a measurement of the $M_X$ spectrum at Belle-II.

\section{Distribution Functions in $B \rightarrow X_u \ell \bar{\nu}_{\ell}$}
Our starting point is the triple differential distribution for $B\to X_u \ell\nu$, which in the case of a massless lepton can be written as
\begin{eqnarray} 
\frac{d^3 \Gamma}{d\qq \,d\qz \,d\el} &=&
\frac{G_F^2 |V_{ub}|^2}{8\pi^3}
 \Bigl\{ 
\qq W_1- \!\left[ 2\el^2-2\qz \el + \frac{\qq}{2} \right] \!W_2 
+ \qq (2\el-\qz) W_3 \Bigr\} 
\nonumber\\
&& \hspace{2cm}
\times
\, \theta \left(\qz-\el-\frac{\qq}{4\el} \right) 
 \theta(\el) \ \theta(\qq)\ \theta(\qz-\sqrt{\qq}),\label{eq:aquila_normalization}
\end{eqnarray}
where $q_0$ and $E_\ell$ are the total leptonic and the charged lepton energies
in the $B$ meson rest frame and $q^2$ is the leptonic invariant mass. The three structure functions $W_i(q_0,q^2)$ are in turn given by the convolution \cite{Gambino:2007rp}
\begin{equation} \label{eq:conv2}
 W_i(q_0,q^2) = m_b^{n_i}(\mu)
 \int F_i(k_+,q^2,\mu) \ W_i^{pert}
\left[ q_0 - \frac{k_+}{2} \left( 1 - \frac{q^2}{m_b M_B} \right), q^2,\mu \right] dk_+
\end{equation}
where $n_{1,2}=-1,n_3=-2$.
The perturbative kernels $W_i^{pert}$ are computed in the kinetic scheme \cite{kin}
with a hard cutoff $\mu$; in the present implementation \cite{Gambino:2007rp}
effects up to $O(\as^2\beta_0)$ are included. 
Eq.~(\ref{eq:conv2}) defines the SFs, $F_i(k_+, q^2, \mu)$, which describe 
the Fermi motion as well as other subleading effects. The $k_+$ moments of the $F_i$
are fixed by the local OPE, which provides $W_i^{pert}$, and by Eq.~(\ref{eq:conv2}). As long as perturbative 
corrections to the Wilson coefficients of the power suppressed operators are neglected,
they are given by 
\be\label{momentsOPE}
\int k_+^n \ F_i(k_+, q^2, \mu) \ dk_+= \left( \frac{2 m_b}{1-\frac{q^2}{m_b M_B}}\right)^n \left[
\delta_{n0} + \frac{I_i^{(n),pow}}{I_i^{(0),tree}}\right],
\ee
where $I_i^{(n)}$ represents the $n$-th central $q_0$ moment of $W_i$, reported in Appendix B of Ref.~\cite{Gambino:2007rp}. All moments but the zero-th one vanish in the limit of infinite $m_b$ and are expressed in terms of the OPE parameters. For illustration, the second moment of $F_3$ is given by
\be
\int k_+^2 F_3(k_+, q^2, \mu) dk_+=\left( \frac{1}{1-\frac{q^2}{m_b M_B}}\right)^2
\left[ (1-\y)^2 \left(\frac{\mupi}3 - \frac{\rls}{3m_b}\right) -(1-\y)\y \frac{2\rd}{3m_b} \right],\label{I2f3}
\ee
up to $ O(\as\Lambda^2/m_b^2, \Lambda^4/m_b^4)$ corrections. Here $\y=q^2/m_b^2$; $\mupi,\mug,\rd,\rls$ are the $B$-meson matrix elements 
of the local dimension 5 and 6 operators that appear in the local OPE and are known 
from fits to the moments of semileptonic decays into charm, see \cite{Alberti:2014yda}
 for recent results.
In the kinetic scheme, the cutoff dependence of these matrix elements propagates to
$F_i$ in such a way that (\ref{eq:conv2}) is order by order independent of $\mu$.
In the limit $q^2\to 0, m_b\to \infty$ one recognizes the moment of the leading SF. 
As discussed in \cite{Gambino:2007rp}, the formalism applies only to low and moderate
$q^2$. At high $q^2$ there is no hard scale and the contribution of higher dimensional operators is no longer suppressed. We therefore use Eq.~(\ref{eq:conv2}) only for $q^2<q_*^2=11$GeV$^2$. At higher $q^2$ the rate must be modelled and we employ
the second method described in Sec.~5 of Ref.~\cite{Gambino:2007rp}.
It is worth stressing that the SFs moments typically have a 20-30\% uncertainty, due to missing higher orders in the OPE, and to the limited precision with which the OPE parameters are known.

Since most of the available information on the distribution functions $F_i$ concerns their first two moments, one option is to assume for them a two-parameter functional form,
such as the exponential 
\be\label{basic1}
F(k_+)= N \,(\bar{\Lambda}-k_+)^a \, e^{b \,k_+}\
 \theta(\bar\Lambda-k_+),\nonumber
\ee
 determining the normalization $N$ and the parameters $a$, and $b$ from the moments. An extensive set of two-parameter functional forms has been considered in \cite{Gambino:2007rp}, with the two parameters and the normalization determined in bins of $q^2$. 
Even though the variation in $|V_{ub}|$ due to the choice of functional form within this set appears rather 
small (typically 1-2\%), this method has obvious intrinsic limitations
and lacks the flexibility to adapt to new experimental 
information which should become available at Belle-II.
In this paper we explore a different path, training neural networks as functions 
of $k_+$ and $q^2$ on the moments.
In the future,
the training will involve also experimentally measured distributions.
The training yields neural network replicas which correspond to analytic parameterizations 
of the functions $F_i$; they can be employed in Eqs.~(\ref{eq:aquila_normalization},\ref{eq:conv2}) to compute the branching fraction subject to given experimental cuts and, comparing this with its experimental measurement, to extract $|V_{ub}|$. 

After the calculation of the complete $O(\as^2)$ and $O(\as \Lambda^2/m_b^2)$ corrections to the moments and rate of $B\to X_u \ell\nu$ \cite{Brucherseifer:2013cu,Alberti:2013kxa,Mannel:2015jka}, the implementation of the GGOU approach described in Ref.~\cite{Gambino:2007rp} needs to be updated. While we do not expect large effects on the extraction of $|V_{ub}|$, we leave this task to a future publication. 

\section{Artificial Neural Networks} 
Artificial Neural Networks (NNs) provide an unbiased parameterization of a continuous function. They consist of a nonlinear map between a space of inputs and a space of outputs, and are {\it universal approximators}, in the sense that they can approximate any continuous function with arbitrary accuracy, provided that sufficiently many nodes are available (for the case of feedforward NN, see \cite{hornik}).  
Finite-size networks are limited in accuracy, but unlike the truncated expansion of a function in a complete basis, their nonlinear nature ensures that this is not a source of bias, as can also be checked by increasing the size of the network. For an elementary introduction to NNs, see \cite{book}. NNs have been successfully employed in many applications in high energy physics, {\it e.g}.\ in the parameterization of PDFs by the NNPDF Collaboration \cite{Ball:2008by}, and in countless experimental analyses, from tagging to triggering.


\subsection{Structure of a Feed-Forward Neural Network}
A simple Feed-Forward Artificial NN is a tuneable analytic sequence of operations on an array of input values in an attempt to recreate a desired output. The most basic system is a single-node (neuron) where a pair of inputs are weighted by adjustable multiplicative parameters, and the sum is then fed into an activation function to produce an output. By changing these weights one can adjust their network to mimic a desired operation. By combining multiple layers of nodes more complex outputs and functions can be obtained.
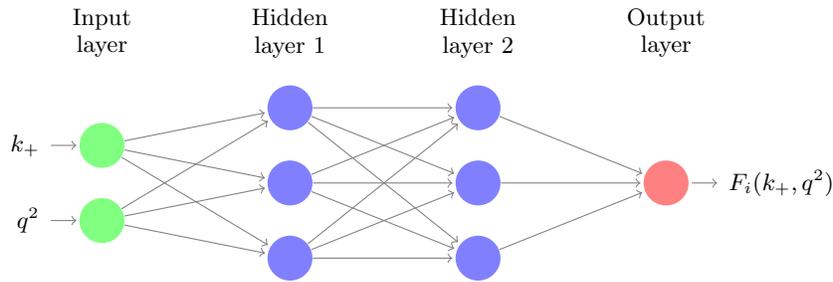
\begin{figure}[t]
\begin{center}
	\def\layersep{2.5cm}
	\begin{tikzpicture}[shorten >=1pt,->,draw=black!50, node distance=\layersep]
	\tikzstyle{every pin edge}=[<-,shorten <=1pt]
	\tikzstyle{neuron}=[circle,fill=black!25,minimum size=17pt,inner sep=0pt]
	\tikzstyle{input neuron}=[neuron, fill=green!50];
	\tikzstyle{output neuron}=[neuron, fill=red!50];
	\tikzstyle{hidden neuron}=[neuron, fill=blue!50];
	\tikzstyle{annot} = [text width=4em, text centered]
	
	\node[input neuron, pin=left:$k_+$] (I-1) at (0,-1) {};	
	\node[input neuron, pin=left:$q^2$] (I-2) at (0,-2) {};	
	
	\foreach \name / \y in {1,...,3}
	\path[yshift=0.5cm]
	node[hidden neuron] (H-\name) at (\layersep,-\y + 0) {};
	
	\foreach \name / \y in {1,...,3}
	\path[yshift=0.5cm]
	node[hidden neuron] (HH-\name) at (\layersep + \layersep , -\y ) {};
	
	\node[output neuron,pin={[pin edge={->}]right:$F_i(k_+,q^2)$}, right of=HH-2] (O) {};
	
	\foreach \source in {1,...,2}
	\foreach \dest in {1,...,3}
	\path (I-\source) edge (H-\dest);
	
	\foreach \source in {1,...,3}
	\foreach \dest in {1,...,3}
	\path (H-\source) edge (HH-\dest);
	
	\foreach \source in {1,...,3}
	\path (HH-\source) edge (O);
	
	\node[annot,above of=H-1, node distance=1cm] (hl) {Hidden layer 1};
	\node[annot,above of=HH-1, node distance=1cm] (hll) {Hidden layer 2};
	\node[annot,left of=hl] {Input layer};
	\node[annot,right of=hll] {Output layer};
	\end{tikzpicture}
\vspace{0.2cm}
\end{center}
\caption{Neural Network with $\{2,3,3,1\}$ architecture.}\label{fig1}
\end{figure}
The notation used to define the initial and subsequent NN structures will be described by their node layout, i.e. $\mathbf{ \{2,3,3,1\} }$, see Fig.~\ref{fig1}. This represents a NN with 2 inputs, 1 output, and two sequential hidden layers with nodes each. The 
inputs $\xi_j^{(l-1)}$ to the node $i$ in layer $l$ are combined into
\be
\xi_i^{(l)}=g^{(l)}\left( \sum_{j=1}^{n_{l-1}} w_{ij}^{(l-1)}\xi_j^{(l-1)}-\theta^{l}_i\right),
\ee
where $w_{ij}^{(l-1)}$ are the weights of the connections leading to this particular node and $\theta^{l}_i$ a {\it threshold} which is trained along with the weights. While our standard includes in total 7 hidden nodes, it can be advantageous to increase the system size to ensure convergence, as will be noted further. The number of parameters in a network depends on the number of nodes per layer, $n_\ell$, and for one or more hidden layers is
\bea
N_p &= &(n_0 + 1) n_1 + (n_1 + 1)n_2 + (n_2 + 1) n_3 + \dots \nonumber
\eea
For example a structure $\mathbf{ \{2,3,3,1\} }$ has $25$ parameters. We have tested one and two layers and eventually employed the architecture $\mathbf{ \{2,7,1\} }$.

Various choices are possible for the activation function $g^{(l)}$, including $g(x)=\tanh x$, $g(x)= x/(1+|x|)$, and the sigmoid function
\begin{equation}
g(x) = \frac1{1+e^{-x}}.
\end{equation}
We generally employ $\tanh$ for the hidden nodes and the sigmoid for the final, to ensure positivity of the output. Changing the activation function in principle should only affect the training time and can be catered to each specific problem. For example, when convolutions of the SFs are required at each training step it is beneficial to switch to a network that employs $g(x)=x/(1+|x|)$, as the performance boost is significant. The two inputs of our NN correspond to the arguments $k_{+}$ and $q^2$ of the SFs $F_i$, both re-scaled to the interval $ (0,1)$.

\subsection{Basic Genetic Algorithm}
The network is trained using a basic Genetic Algorithm, whereupon each child generation is created from a single parent through a series of randomly selected variations on the weights of the NN. The top children as determined through an error analysis of the output are kept and they become the new parents. Each parent-child generation will be referred to as an epoch.

We begin training the NN with randomized weights. The number of variations to be made is randomly selected between 1 and 3. Each selected weight $w_{ji}$ is modified by
\begin{equation}
w'_{ji} = w_{ji} + r_i \times \eta_{NN}(g_e),
\end{equation}
where $r_i$ is a random number from $-1$ to $1$ for each weight; $\eta_{NN}(g_e)$ is a learning rate that starts at $5.0$ initially, but adaptively adjusts depending on the activity of the network; and $g_e$ is the current global epoch number. With this method large variations that would not be beneficial occur less frequently as the epochs pass. If there is no activity for a certain amount of time there is a chance that a local minimum has been found, and $\eta_{NN}$ begins increasing to allow for solutions that can escape this minimum. The learning rate and method with which the weights are adjusted should only affect the learning efficiency and should not introduce a bias in the final replica results. 

The process above is repeated 20 independent times on the parent network, and the best resulting child becomes the new parent for the following epoch. The learning criterion, or ``Goodness of Fit'' measurement, is defined by the user. For different cases one can use different requirements for training. We choose to use multiple methods, which are detailed in the following section, in order to gauge the validity of this approach.
\begin{figure}[t]
\begin{center}
\includegraphics[width=11.5cm]{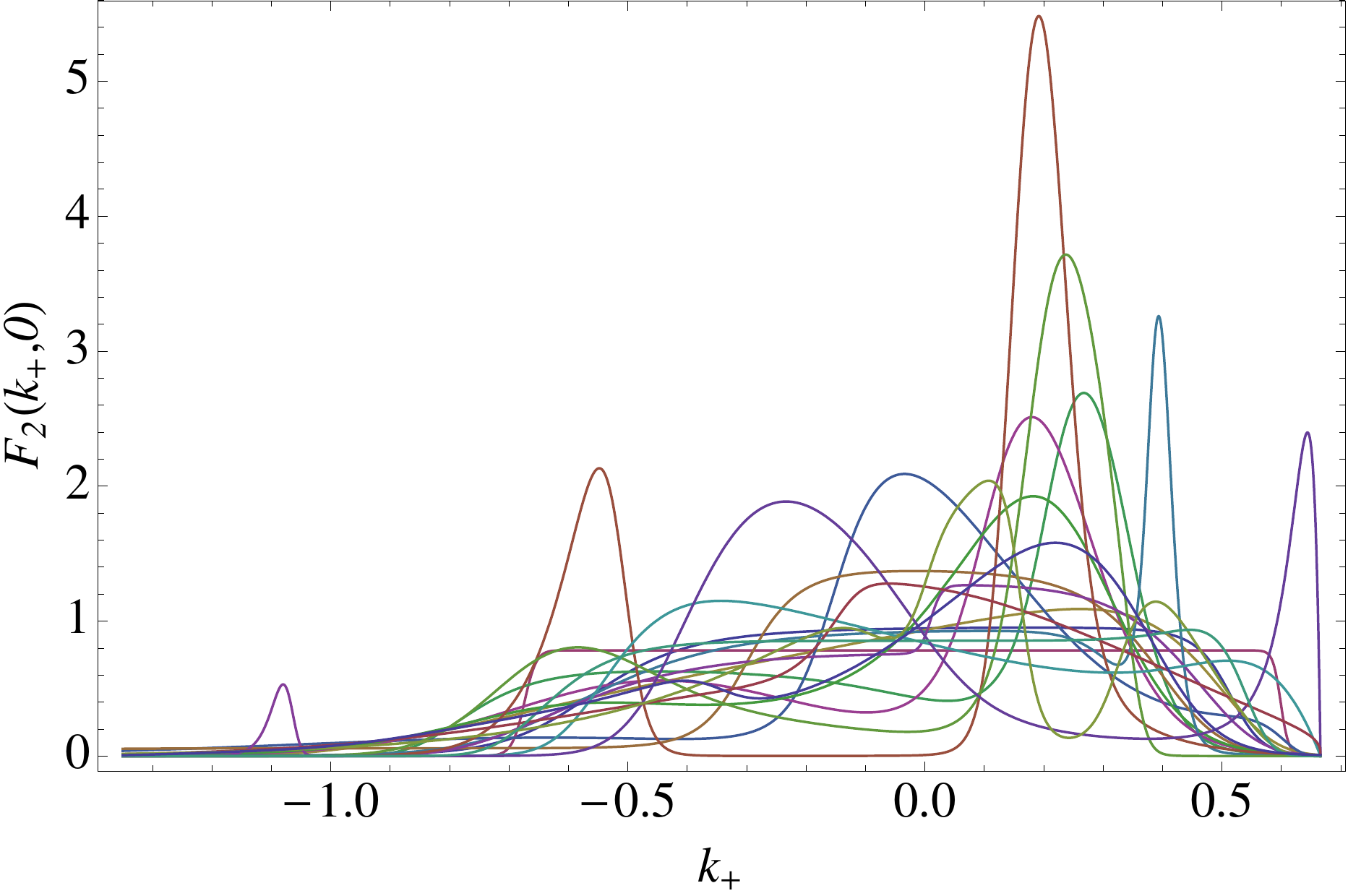} 
\end{center}
\caption{Selection of NN replicas of $F_2(k_+,0)$ trained on the first three moments only.}
\label{NNsample}
\end{figure}

\subsection{Error Minimization}
The NN training is governed by an error function which in our case is the $\chi^2$
obtained by comparing the calculation of the first 
$k_{+}$-moments of a given SF for a selected set of $q^2 \in [0\, \mbox{GeV}^2, 13\, \mbox{GeV}^2]$ with the OPE 
constraints. Initially we use 7 evenly spaced 
 values of $q^2$. We can select an alternate set of $q^2$ values in the same range to test against over-fitting the sampling points in the $q^2$ direction, but we have verified 
 that this is not an issue and the functions remain smooth in $q^2$. For each value of $q^2$ we compute the normalization and first three moments in $k_{+}$, 
\begin{equation}
M_{n,i}(q^2) = \int_{-\infty}^{\bar\Lambda}k_{+}^n F_i(k_{+},q^2) \,dk_{+} ,
\end{equation}
where $M_{n,i}(q^2)$ are given in Eq.~(\ref{momentsOPE}). 
There are therefore 28 quantities to be fit.
However, we generally employ the third moment only as a loose constraint, 
and the normalization of the second shape function is fixed to 1 at the order we are working. Throughout the learning phase we monitor the evolution of the $\chi^2$, computed in the various cases as detailed below. The scarcity of data makes it impossible to use a control sample, as done by the NNPDF collaboration. The $\chi^2$ first decreases quickly, with training progressively slowing as expected. We stop the learning when a certain condition is met, typically when 
the $\chi^2$ of each replica reaches a certain value.

It is worth stressing that
 the first two or three moments do not constrain the SFs much. The point is illustrated
 in Fig.~\ref{NNsample} by a representative selection of NN for $F_2(k_+, 0)$, which 
are normalized to 1 and satisfy the first two moments within a few \% and 
 and the third moment within 60\%. 
A tighter constraint on the third moment would not change this picture significantly.
Of course, not all the shapes shown in this plot are physically acceptable and 
only a handful of them can roughly reproduce the photon spectrum in $B\to X_s \gamma$.
However, this plot demonstrates the capability of NN to properly sample the functional space. 

One should be aware that the sampling can be biased in several ways, 
for instance by selection based on the speed of learning, by improper choice of random initial weights or by the use of an underlying function to speed the training up.
Indeed, in order to decrease the learning time and to ensure the vanishing of the SFs at the endpoint, we scale the network output by a function that provides the proper behavior. We know the SFs must approach zero at $-\infty$, and cut off at $\bar \Lambda$. To ensure this, one option is to define our full SFs as 
\begin{equation}
F_i(k_{+},q^2) = (c_{i0}+ c_{i1} q^2) \,e^{(c_{i2}+c_{i3} q^2) k_{+}} \,(\bar\Lambda - k_{+})^{(c_{i4}+c_{i5} q^2)} \,N_i(k_{+},q^2),
\end{equation}
where $N_i$ is the NN function to be trained. The coefficients $c_{ij}$, are trained simultaneously with the NN weights and are unconstrained. In the case of 
the $\mathbf{ \{2,7,1\} }$ architecture, which we generally adopt below, we therefore have a total of 35 parameters.
In order to minimize the bias we have used a set of different underlying functions, although there would be no bias if the SFs were sufficiently constrained by experimental data. 
 \begin{figure}[t]
\begin{center}
\includegraphics[width=17cm]{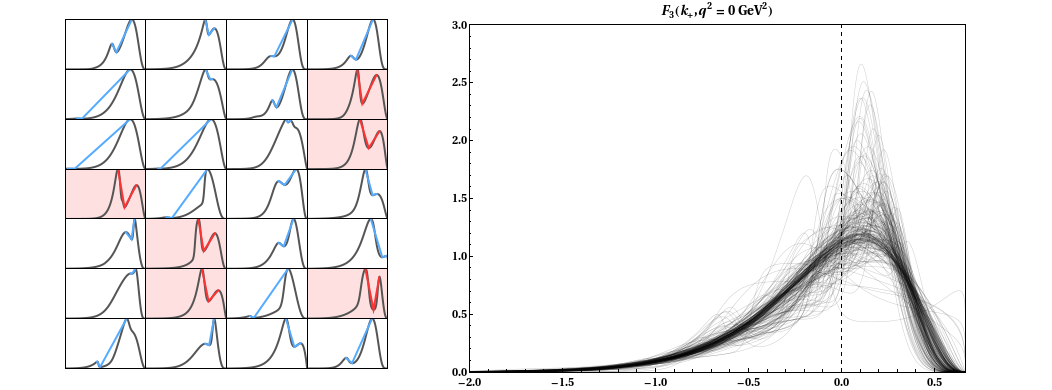} 
\end{center}
\caption{Left: examples of accepted and rejected (in red) shapes. Right: sample of NN replicas of $F_3(k_+,0)$ trained on the first three moments only after applying the selection criteria.}
\label{NNsample2}
\end{figure}

As already mentioned, additional information on the SFs comes from the photon spectrum measured in inclusive radiative B decays. One could include these data with 
an additional $O(10\%)$ theoretical uncertainty to account for power suppressed corrections to the relation between the photon and semileptonic SFs at $q^2=0$. We 
postpone a careful study of the photon spectrum to a future publication. 
However, in the present pilot study we  include the main qualitative features of the experimental photon spectrum, assuming that the SFs are all dominated by a single peak (without excluding multiple peaks) and are never too steep. As we will illustrate in a moment, these minimal assumptions 
strongly reduce the variety of functional forms, as would also do a measurement of the $M_X
$ spectrum at Belle-II.
 
\section{Results and discussion}

{\bf A. } \ As a first step, we train the NN on the moments only and compare with the functional form error found in \cite{Gambino:2007rp}. 
At this stage we are only interested in the spread of the replicas in functional space. 
To this end we compute the moments with the same (outdated) input parameters used in \cite{Gambino:2007rp}, neglecting all uncertainties and correlations. Each NN replica is required to reproduce the moments at seven equally spaced 
$q^2$ points between $q^2=0$ and 13GeV$^2$. 
The training is stopped when 
 $\chi^2= n$, where $n$ is the total number of constraints, and $\chi^2$ is computed using relative errors of 3\% on the normalization, first and second moment, and of 10\% on the third moment, assuming no correlation between different moments and different bins in $q^2$.
 The training is rather long and becomes very slow for smaller errors. 
 After training a sample of 
 NNs we select those whose derivative never exceeds 50 in absolute value and which have only one dominant peak (in the case of multiple peaks we check that the height of the subdominant ones is less than 20\% of the height of the dominant one, measured wrt the 
 common trough). A representative sample of accepted and rejected shapes is shown on the 
 left in Fig.~\ref{NNsample2}, while   on the right  we display a
 sample of about 150 replicas for $F_3(k_+,0)$  after this pruning.

\begin{figure}[t]
\begin{center}
\includegraphics[width=12cm]{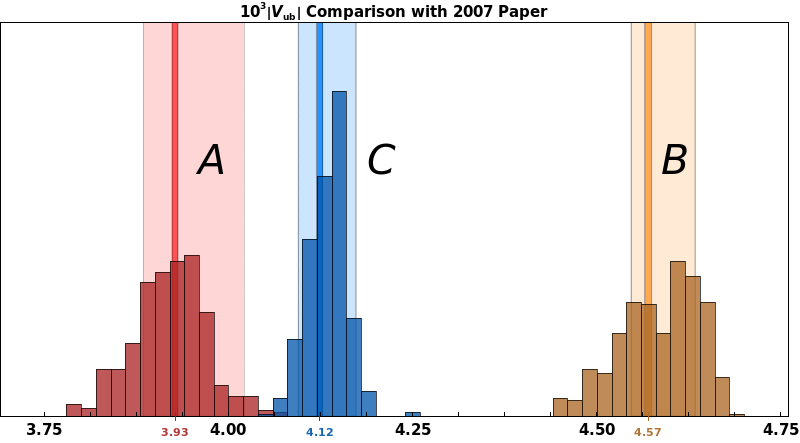}
\end{center}
\caption{Comparison between the $|V_{ub}|$ ranges due to functional form variations given in \cite{Gambino:2007rp} and the values obtained using NN trained on moments computed with the same inputs used there.}
\label{VubA}
\end{figure}

Each triplet of the selected NN replicas of $F_{1-3}(k_+,q^2)$ then allows for a determination of $|V_{ub}|$ when it is confronted with the 
experimental results for a given partial BR. In order
to compare with the results given in the GGOU original paper we compute $|V_{ub}|$
from the same four specific experimental results used there,
namely
\begin{enumerate}
\item [A] $M_X$ cut: $M_X < 1.7, E_{\ell} > 1.0 $ GeV, Belle \cite{Bizjak:2005hn};
\item [B] Combined $M_X$ and $q^2$ cuts: $M_X \leq 1.7 {\rm GeV},\ q^2 > 8 {\rm GeV}^2, E_{\ell} > 1.0$ GeV, Babar \& Belle \cite{Bizjak:2005hn,Aubert:2005hb};
\item [C] Lepton endpoint: $E_{\ell} > 2.0$ GeV, Babar \cite{Aubert:2005mg},
\end{enumerate}
and compare the spread in $|V_{ub}|$ with the functional form dependence 
given in \cite{Gambino:2007rp}. This is illustrated in Fig.~\ref{VubA}, where
the spread in the value of $|V_{ub}|$ measures the SFs uncertainty.
We have checked that using different NN architectures
leads to very similar results. 
In the calculation of the partial rates we use the same high-$q^2$ setting used by \cite{Gambino:2007rp} for the functional form uncertainty, namely the second method described in Sec.~5 of that paper.
We observe that the central values are very close to those obtained in \cite{Gambino:2007rp}. The spread in the 
$|V_{ub}|$ values is larger than in 2007, but the standard deviation of the distributions are
roughly comparable with the functional form errors found in that analysis.

\vspace{2mm}

{ \bf B.} \ 
As a second step, we include in the analysis the complete theoretical and parametric uncertainty on the moments, with all the correlations between moments and different 
$q^2$ bins. Here we want to show that the method allows us to include multiple data with 
non-trivial correlations and that the errors and correlations in the inputs are reproduced by 
the ensemble of trained replicas.
The OPE parameters are taken from \cite{Alberti:2014yda} and the theoretical uncertainties 
of the $F_i$ moments are estimated as in that paper. The theoretical correlation between different $q^2$ bins is 
estimated with method C in Sec. 3 of \cite{Gambino:2013rza}. After adding the  covariance matrices related to the input parameters and to the theoretical uncertainties, a replica of  pseudo-data for the moments of 
the three SFs is produced assuming 
gaussian distributions. The NN  for each $F_i$ are then trained on this replica, keeping track of the 
input parameters, and in particular of $m_b$, which is used in the calculation of physical quantities from Eq.\,(\ref{eq:aquila_normalization}).
The training is again ruled by the $\chi^2$ function, which now includes all correlations.
Even though the typical total uncertainty of the first three moments is as large as 25-30\%, high correlations between $q^2$ bins do not allow to speed up the training significantly.

\begin{table}[t]
\begin{center}
\begin{tabular}{|l|c|c|}
\hline Experimental cuts (in GeV or GeV$^2$) & $|V_{ub}| \times 10^{3}$ & $|V_{ub}| \times 10^{3}$\cite{hfag}\\ \hline
 $M_X < 1.55, E_{\ell} > 1.0 $ Babar \cite{babarnew}& $4.30(20)(^{26}_{27})$ &$4.29(20)(^{21}_{22}) $ \\
 $M_X < 1.7, E_{\ell} > 1.0 $ Babar \cite{babarnew}& $4.05(23)(^{19}_{20})$ & $4.09(23)(^{18}_{19}) $ \\
 $M_X \leq 1.7 , q^2 > 8, E_{\ell} > 1.0$ Babar\cite{babarnew}& $4.23(23)(^{26}_{28})$ &$4.32(23)(^{27}_{30}) $ \\
 $E_{\ell} > 2.0$ Babar \cite{Aubert:2005mg}& $4.47(26)(^{22}_{27})$ &$4.50(26)(^{18}_{25}) $ \\ 
 $E_{\ell} > 1.0$ Belle \cite{Urquijo:2009tp}&$4.58(27)(^{10}_{11})$ &$4.60(27)(^{10}_{11}) $ \\ 
 \hline
\end{tabular}
\end{center}
\caption{$|V_{ub}|$ determinations using different experimental analyses and comparison with HFAG latest results in the GGOU approach \cite{hfag}. }
\label{table1}
\end{table}

As we adopt up-to-date inputs, we can extract $|V_{ub}|$ from 
the latest experimental results and compare the results with the most recent HFAG 
compilation \cite{hfag}; this is done in Table 1 for the most representative cases, using 
the isospin average $\tau_B=1.582(5)$ps, employed in the derivation of the HFAG values.
 The first uncertainty represents the total experimental error, while the second is the sum in quadrature of   the  standard deviation of the values obtained by the replicas (which 
in this  case accounts for both functional form and parametric uncertainties), 
and  all the remaining theoretical uncertainties (perturbative, treatment of the high $q^2$ tail, Weak Annihilation), which are estimated in the same way as in \cite{Gambino:2007rp}. 

While we refrain from combining the values of $|V_{ub}|$ originating from different experimental analyses,
we observe that the central values are quite close to those obtained by HFAG. A minor shift downwards is to be expected because, following \cite{Alberti:2014yda},   we adopt a slightly 
higher $m_b$ than employed by HFAG.
The theoretical errors, which are asymmetric because of the Weak Annihilation error, are generally slightly larger than those reported by HFAG, especially when the cuts make $|V_{ub}|$ more sensitive to the SFs. This is due to $i)$ a larger spread in 
the functional space of the $F_i$ compared with the  method of Ref.~\cite{Gambino:2007rp}
used by HFAG; $ii)$ the introduction of a non-negligible theoretical error for the SFs moments, which was not considered in \cite{Gambino:2007rp}.

Given the uncertainties, the agreement between the different rows of Table 1 is good,
and shows that the OPE based approach describes the present data on $B\to X_u \ell \nu$ reasonably well.  
We also notice that the HFAG average, Eq.~(1), is dominated by the Belle analysis 
\cite{Urquijo:2009tp}, reported in the last row of Table 1, 
and by a similar Babar analysis with a $p^*>1.3$GeV cut, since they have a significantly 
smaller theoretical error and both prefer a high $|V_{ub}|$. 
However, as already mentioned in the Introduction, these analyses heavily depend on  background subtraction and signal modelling. On the other hand, 
the values reported in the first three rows of Table 1 are consistent with 
the recent exclusive $|V_{ub}|$ results given in \cite{Lattice:2015tia} and \cite{Flynn:2015mha}  within 1.5$\sigma$.
\vspace{2mm}

\begin{figure}[t]
\begin{center}
\includegraphics[width=12cm]{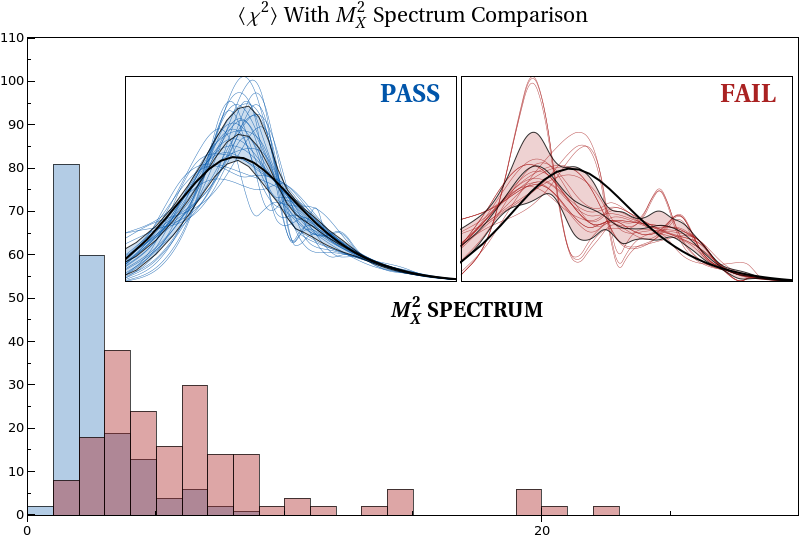}
\end{center}
\caption{Sample $\chi^2_X$ comparison of the step-A sample of NNs with simulated $M_X^2$ spectra data. Inset are the 2-peak pruning analysis and  the resulting $M_X^2$ spectra for the pass(blue)/fail(red) ensembles, with the reference spectrum in black.}
\label{Spectra_Pruned}
\end{figure}

{ \bf C.} \ Finally, we consider the possible impact of future Belle-II data on the SFs and consequently on the $|V_{ub}|$ determination. We assume that 
Belle-II will measure the $M_X$ spectrum in 8 evenly spaced bins below $M_X=2$GeV, with a total 4\% uncertainty in each bin. A detailed estimate of the potential improvement in 
the $|V_{ub}|$ determination would involve a lengthy training of the NN on both the moments and these new data and is  beyond the scope of this paper. 
Here we demonstrate the discriminating power of the $M_X$ spectrum data   
using the NN replicas obtained in step A above, all of which  reproduce precisely the first moments. We use randomly selected triplets of these NNs to compute the $M_X$-spectrum
and compare the results with a reference $M_X$-spectrum obtained using one of these triplets. The results are shown in the two insets in Fig.\,\ref{Spectra_Pruned}.
The plot on the lhs refers to replicas which survived the pruning described in {\bf A},
while on the rhs the spectra are produced based on replicas that have been trained on the moments, but failed our acceptance criteria. In both cases the shaded band corresponds to the 1$\sigma$ band around the central value. We observe that the $M_X$-spectrum
is very sensitive to the presence of sharp features in the SFs, which are more likely in the 
rejected sample, and that a precise measurement of the spectrum can even exclude many of the replicas in the accepted sample. The Belle-II data therefore have the potential to 
constrain significantly the SFs and to validate the OPE-based approach to inclusive charmless 
semileptonic $B$ decays.

The above considerations can be made more quantitative by defining a $\chi^2_X$ based on the comparison of the yield in each $M_X$-bin computed from a given triplet with the yield 
computed from the reference  (``simulated'') $M_X$-spectrum data, assuming a total 4\% error.
The main graph in Fig.\,\ref{Spectra_Pruned} shows the distributions of accepted and rejected replicas as a function of $\chi^2_X$. Most of the rejected replicas and many of the accepted ones would be excluded 
by a test based on $\chi^2_X$. Indeed, one can {\it reweight} the NN replicas using
their $\chi^2_X$, giving more importance to the replicas whose $M_X$ spectrum is closer to the experimental one and therefore have lower $\chi^2_X$, see \cite{Ball:2010gb}.
Performing such an exercise on our step A shows that reweighting with the $\chi^2_X$ reduces the uncertainty from the functional forms by 30-70\%, depending on the experimental cuts, and induces a $(0.2-0.4)\%$ negative shift in $|V_{ub}|$.

Of course, the $M_X$-spectrum carries information not only on the SFs, but also on the
HQE parameters $m_b$, $\mupi$, etc.\ which have been fixed in the exercise we have just discussed. This is related to what one can learn from the moments of the
$M_X$ spectrum, see \cite{Gambino:2005tp} for a discussion of their sensitivity to the HQE parameters. As a consequence,  reweighting  the replicas of step B based on the same data would have a much more dramatic effect, because their first moments have a much larger spread.
  Unfortunately, in step B the number of available replicas is too limited. The main point to be emphasised, however, is that in our framework the $B\to X_u \ell\nu$ kinematic distributions
($M_X, q^2, E_\ell$ spectra)   can be considered together
with all the available relevant information ($B\to X_c \ell\nu$ moments,  $B\to X_
\gamma$ spectrum, $m_{b,c}$ determinations, etc.), in the context of a NN training where the HQE parameters are fitted together with the NN parameters.  Such analysis 
will be mandatory with Belle-II data. 

\section{Summary}
We have introduced a new parameterization of the SFs characterizing  inclusive semileptonic $B$ decays without charm based on artificial neural networks. 
The new method  allows 
for alternative, unbiased estimates of the SFs functional form uncertainty, which turn out 
to be in reasonable agreement with previous results obtained  using
functional form models. 
 As we have shown explicitly, a clear advantage of the method is that  it permits a straightforward implementation  of new experimental data, such as those which will become available at Belle-II.  These data will 
reduce the SFs uncertainty and, most importantly, their comparison with high-precision theoretical predictions will validate the OPE-based approach in a much more stringent way.

\begin{acknowledgments}
We are grateful to Alberto Guffanti and Hannes Zechlin for very useful discussions. We thank the Mainz Institute for Theoretical Physics (MITP) for hospitality and partial support during the workshop {\it Challenges in semileptonic B decays} in April 2015, where part 
of the work was done. Work supported in part by MIUR under contract 2010YJ2NYW 006.

\end{acknowledgments}


\end{document}